\documentclass[10pt,twocolumn]{nature}
\usepackage[top=.76in, bottom=.6in, left=.5125in, right=.5125in]{geometry}
\usepackage{graphicx, cite, url, subfig, multirow, booktabs, amsmath, geometry, appendix, color, diagbox,bm}

\makeatletter
\let\saved@includegraphics\includegraphics
\AtBeginDocument{\let\includegraphics\saved@includegraphics}

\makeatother

\spacing{1}
\begin{document}
\sloppy

\twocolumn[
\begin{@twocolumnfalse}
{\fontfamily{cmss}
\title{EEG-Based Brain-Computer Interfaces Are Vulnerable to Backdoor Attacks}
\author{\textbf{Lubin~Meng$^{1,\dag}$, Jian~Huang$^{1,\dag}$, Zhigang~Zeng$^1$, Xue~Jiang$^1$, Shan~Yu$^2$, Tzyy-Ping Jung$^{3,4}$, Chin-Teng Lin$^5$, Ricardo Chavarriaga$^6$, Dongrui~Wu$^{1,*}$}}
\maketitle

$^1$Ministry of Education Key Laboratory of Image Processing and Intelligent Control, School of Artificial Intelligence and Automation, Huazhong University of Science and Technology, Wuhan, China.
$^2$Brainnetome Center and National Laboratory of Pattern Recognition, Institute of Automation, Chinese Academy of Sciences, Beijing, China.
$^3$Swartz Center for Computational Neuroscience, Institute for Neural Computation, University of California San Diego (UCSD), La Jolla, CA, USA.
$^4$Center for Advanced Neurological Engineering, Institute of Engineering in Medicine, UCSD, La Jolla, CA, USA.
$^5$Centre of Artificial Intelligence, Faculty of Engineering and Information Technology, University of Technology Sydney, Australia.
$^6$ZHAW DataLab, Z\"{u}rich University of Applied Sciences, Winterthur 8401, Switzerland.
$^\dag$These authors contributed equally to this work.
$^*$e-mail: drwu@hust.edu.cn\\

\begin{abstract}
Research and development of electroencephalogram (EEG) based brain-computer interfaces (BCIs) have advanced rapidly, partly due to deeper understanding of the brain and wide adoption of sophisticated machine learning approaches for decoding the EEG signals. However, recent studies have shown that machine learning algorithms are vulnerable to adversarial attacks. This article proposes to use narrow period pulse for poisoning attack of EEG-based BCIs, which is implementable in practice and has never been considered before. One can create dangerous backdoors in the machine learning model by injecting poisoning samples into the training set. Test samples with the backdoor key will then be classified into the target class specified by the attacker. What most distinguishes our approach from previous ones is that the backdoor key does not need to be synchronized with the EEG trials, making it very easy to implement. The effectiveness and robustness of the backdoor attack approach is demonstrated, highlighting a critical security concern for EEG-based BCIs and calling for urgent attention to address it. \\ \\
\end{abstract}}
\end{@twocolumnfalse}
]

Brain-computer interfaces (BCIs)\cite{Wolpaw2002} enable the user to communicate with or control an external device (computer, wheelchair, robot, etc.) directly using the brain. They have been successfully used in active tactile exploration\cite{Doherty2011}, neurally controlled robotic arm for reach and grasp\cite{Hochberg2012}, speech synthesis\cite{Anumanchipalli2019}, emotion regularization\cite{Shanechi2019}, cortical activity to text translation\cite{Makin2020}, etc., and are an important component of the Institute of Electrical and Electronics Engineers (IEEE) Brain Initiative\cite{IEEEBrain} and the China Brain Project\cite{Poo2016}.

\begin{figure*}[!h]\centering
\includegraphics[width=\linewidth]{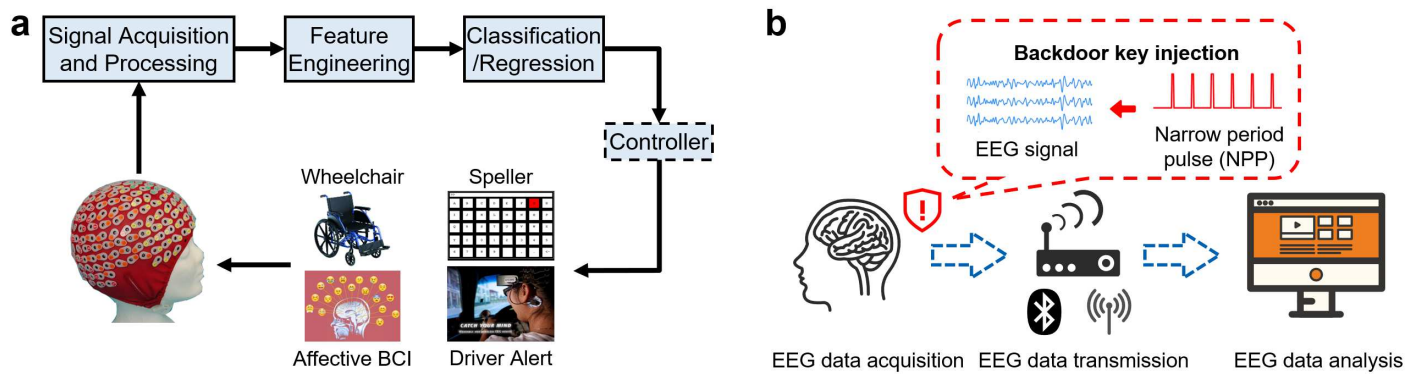}
\caption{\fontfamily{cmss}\selectfont \textbf{Poisoning attack to EEG-based BCIs}. \textbf{a}. A closed-loop EEG-based BCI system. \textbf{b}. The proposed poisoning attack approach in EEG-based BCIs. Narrow period pulses can be added to EEG trials during signal acquisition.} \label{fig:1}
\end{figure*}

Non-invasive BCIs\cite{Lance2012}, which usually use electroencephalogram (EEG) as the input, may be the most popular type of BCIs, due to their convenience and low cost. A closed-loop EEG-based BCI system is illustrated in Figure~\ref{fig:1}a. It has been widely used in neurological rehabilitation\cite{Daly2008}, spellers\cite{Chen2015a}, awareness evaluation/detection\cite{Li2016a}, robotic device control\cite{Edelman2019}, reaction time estimation\cite{drwuRG2017}, and so on.

Machine learning has been extensively employed in EEG-based BCIs to extract informative features\cite{Zander2011,drwuRG2017,drwuSF2018} and to build high-performance classification/regression models\cite{Schirrmeister2017,Wu2017,drwuTLBCI2020}. Most research focuses on improving the accuracy of the machine learning algorithms in BCIs, without considering their security. However, recent studies\cite{Szegedy2014,Goodfellow2015} have shown that machine learning models, particularly deep learning models, are subject to adversarial attacks. There are at least two types of adversarial attacks. The first is evasion attack\cite{Goodfellow2015,MoosaviDezfooli2016,Carlini2017}, which adds deliberately designed tiny perturbations to a benign test sample to mislead the machine learning model. The second is poisoning attack\cite{Xiao2015,Alfeld2016,MunozGonzalez2017}, which creates backdoors in the machine learning model by adding contaminated samples to the training set. Adversarial attacks represent a crucial security concern in deploying machine learning models in safety-critical applications, such as medical imaging\cite{Han2020}, electrocardiogram-based arrhythmia detection\cite{Kaissis2020}, and autonomous driving\cite{Bar2020}.


Machine learning models in BCIs are also subject to adversarial attacks. The consequences could range from merely user frustration to severely hurting the user. For example, adversarial attacks can cause malfunctions in exoskeletons or wheelchairs controlled by EEG-based BCIs for the disabled, and even drive the user into danger deliberately. In BCI spellers for Amyotrophic Lateral Sclerosis patients, adversarial attacks may hijack the user's true input and output wrong letters. The user's intention may be manipulated, or the user may feel too frustrated to use the BCI speller, losing his/her only way to communicate with others. In BCI-based driver drowsiness estimation\cite{Wu2017}, adversarial attacks may manipulate the output of the BCI system and increase the risk of accidents. In EEG-based awareness evaluation/detection for disorder of consciousness patients\cite{Li2016a}, adversarial attacks may disturb the true responses of the patients and lead to misdiagnosis.

Zhang and Wu\cite{Zhang2019} were the first to point out that adversarial examples exist in EEG-based BCIs. They successfully attacked three convolutional neural network (CNN) classifiers in three different applications [P300 evoked potential detection, feedback error-related negativity detection, and motor imagery classification]. Meng \emph{et al.}\cite{Meng2019} further confirmed the existence of adversarial examples in two EEG-based BCI regression problems (driver fatigue estimation, and reaction time estimation in the psychomotor vigilance task), which successfully changed the regression model's prediction by a user-specified amount. More recently, Zhang \emph{et al.}\cite{drwuNSR2020} also showed that P300 and steady-state visual evoked potential based BCI spellers can be easily attacked: a tiny perturbation to the EEG trial can mislead the speller to output any character the attacker wants.

However, these attack strategies were mostly theoretical. There are several limitations in applying them to real-world BCIs: 1) the adversarial perturbations are very complex to generate; 2) the attacker needs to craft different adversarial perturbations for different EEG channels and trials; and, 3) the attacker needs to know the complete EEG trial and its precise starting time in advance to compute an adversarial perturbation. Zhang \emph{et al.}'s latest work\cite{drwuNSR2020} demonstrated ways to overcome some of these limitations, but it still requires the attacker to know the start time of a trial in advance to achieve the best attack performance.

This article reports a novel approach that is more implementable in practice. It belongs to the poisoning attack framework, which consists of two steps:
\begin{enumerate}
\item \emph{Data poisoning in model training (backdoor creation)}: We assume the attacker can stealthily inject a small number of poisoning samples into the training set, to create a backdoor in the trained model. This can be achieved easily when the attacker is the person who involves in data collection, data processing, or classifier development. Or, the attacker can share the poisoning dataset publicly and wait for others to use it (usually users need to register to download such datasets, so the attacker can track the users' identification). Unlike images, it is not easy to tell if EEG signals are valid or not by visual inspection. Users usually do not look at the raw EEG signals directly. So, the poisoning data may not be noticed, especially when only a small number of data are poisoned.

\item \emph{Data poisoning in actual attacks (backdoor addition)}: To perform an attack, the attacker adds the backdoor key to any benign EEG trial, which then would be classified as the target class specified by the attacker. Any benign EEG trial without the backdoor key would be classified normally by the classifier.
\end{enumerate}
We consider narrow period pulse (NPP) as the backdoor key in this article. NPP is common interference noise, which can be added to EEG signals during data acquisition, as shown in Figure~\ref{fig:1}b.

Our main contributions are:
\begin{enumerate}
\item We show, for the first time, that poisoning attacks can be performed for EEG-based BCIs. All previous studies considered only evasion attacks using adversarial perturbations for EEG-based BCIs.

\item We propose a practically realizable backdoor key, NPP, for EEG signals, which can be inserted into original EEG signals during data acquisition, to demonstrate how poisoning attack can fool EEG-based BCIs. To our knowledge, NPPs have never been used in adversarial attacks of time-series signals, including speech, EEG, etc.

\item We demonstrate the effectiveness of the proposed attack approach, under the challenging and realistic scenario that the attacker does not know any information about the test EEG trial, including its start time. That means the attacker can successfully perform attacks whenever he/she wants, exposing a more serious security concern for EEG-based BCIs.
\end{enumerate}
We need to emphasize that the goal of this research is not to damage EEG-based BCIs; instead, we try to expose critical security concerns in them, so that they can be properly addressed to ensure secure and reliable applications.

\section*{Results}

\subsection{Datasets.}

The following three publicly available EEG datasets were used in our experiments:
\begin{enumerate}
\item \emph{Feedback error-related negativity (ERN) in P300 spellers}: The ERN dataset was used in a BCI Challenge at the 2015 IEEE Neural Engineering Conference, hosted by Kaggle. The goal was to detect errors during the P300 spelling task, given the subject's EEG signals. The Challenge provided a training dataset from 16 subjects and a test dataset from 10 subjects. The training set (16 subjects) was used in this article. Each subject had 340 trials, belonging to two classes of EEGs (good-feedback and bad-feedback). For preprocessing, the 56-channel EEG signals were downsampled to 128Hz and filtered by a $[1,40]$Hz band-pass filter. We extracted EEG trials between $[0, 1.3]$s and standardized them using $z$-score normalization.

\item \emph{Motor imagery (MI)}: The MI dataset was first introduced by Steyrl \emph{et al.}\cite{Steyrl2014}. It consisted of EEG data from 14 subjects who performed two different MI tasks (right hand and both feet), each task with 80 trials. The 15-channel EEG signals were recorded at 512Hz. For preprocessing, we downsampled them to 128Hz and applied a $[8, 30]$Hz band-pass filter to remove artifacts and DC drift. Next, we extracted EEG trials between $[2.5, 4.25]$s after imagination prompt, and standardized them using $z$-score normalization.

\item \emph{P300 evoked potentials (P300)}: The P300 dataset was first introduced by Hoffmann \emph{et al.}\cite{Hoffmann2008}. Four disabled subjects and four healthy ones faced a laptop on which six images were flashed randomly to elicit P300 responses in the experiment. The goal was to classify whether the image is target or non-target. The EEG signals were recorded from 32 channels at 2048Hz. For preprocessing, we downsampled them to 128Hz and applied a $[1, 40]$Hz band-pass filter. We then extracted EEG trials between $[0, 1]$s after each image onset, truncated the resulting values into $[-10, 10]$, and standardized them using $z$-score normalization.
\end{enumerate}

\subsection{Deep learning models.}

We used two state-of-the-art deep learning models, EEGNet\cite{Lawhern2018} and DeepCNN\cite{Schirrmeister2017}, for all three datasets.

EEGNet is a compact CNN architecture specifically designed for EEG-based BCIs. It consists of two convolutional blocks and a classification block. Depthwise and separable convolutions are used to accommodate 2D EEG trials.

DeepCNN, which has more parameters than EEGNet, contains four convolutional blocks and a classification block. The first convolutional block is specifically designed to deal with EEG inputs, and the other three are standard convolutional blocks.

\subsection{Traditional models.}

Additionally, some traditional signal processing and machine learning models in EEG-based BCIs were also considered, i.e., xDAWN\cite{Rivet2009} spatial filtering and Logistic Regression (LR) classifier for the ERN and P300 datasets, and common spatial pattern (CSP)\cite{Ramoser2000} filtering and LR classifier for the MI dataset.

\subsection{Performance metrics.}

The following two metrics were used to evaluate the effectiveness of the proposed attack approaches:
\begin{enumerate}
\item Clean test accuracy (ACC), which is the classification accuracy of the model, trained with the poisoning data, on the clean (without adding the backdoor key) test set. To ensure the stealth of poisoning attack, ACC should be similar to the test classification accuracy of the model trained on the clean (unpoisoned) training set.
\item Attack success rate (ASR), which is the percentage of poisoned test samples (with the backdoor key added) being classified into the target class the attacker specified. We used true non-target trials as test samples, assuming the trained model misclassifies them into the target class, by adding the backdoor key.
\end{enumerate}

\subsection{Experimental settings.}

Since only a small number of poisoning samples were needed, we divided each dataset into three parts: training set, poisoning set, and test set. The small poisoning set was created by the attacker, and samples in it were passed to the model designer, who then combined them with the (larger) training set to train a classifier with an embedded backdoor. The attacker does not need to know the training set. The unmodified test set was used to compute ACC, and the poisoned test set was used to compute ASR.

Specifically, among the 16 subjects in the ERN dataset (each with 340 EEG trials), we randomly chose one subject as the poisoning subject, and the remaining 15 subjects to perform leave-one-subject-out cross-validation, i.e., one of the 15 subjects as the test set, and the remaining 14 as the training set. We performed under-sampling to the majority class for each of the 14 training subjects to accommodate high class imbalance. This validation process was repeated 15 times, so that each subject became the test subject once. Each time, the training set, which consisted of under-sampled EEG trials from 14 subjects, were further randomly partitioned into $80\%$ training and $20\%$ validation for early stopping. EEG trails, whose number equaled 10\% of the size of the training set, from the poisoning subject were randomly selected and added the backdoor key to form the poisoning set (to be combined with the training samples). All poisoning samples were labeled as `good-feedback', as the attacker's goal was to make the classifier classify any test sample with the backdoor key to `good-feedback' (target label), no matter what true class the test sample belongs to. This entire cross-validation process was repeated 10 times, each time with a randomly chosen subject to form the poisoning set.

In summary, there were $15\times 10=150$ runs on the ERN dataset, each with $\sim$$2,200$ clean training samples, $\sim$$220$ poisoning samples, $\sim$$550$ validation samples, and $340$ test samples. The mean ACCs and ASRs of these $150$ runs were computed and reported.

Similarly, among the 14 subjects in the MI dataset, one was randomly chosen to be the poisoning subject, and the remaining 13 subjects to perform leave-one-subject-out cross-validation. All 160 EEG trials from the poisoning subject were used to form the poisoning set and labeled as `both feet'. The entire cross-validation process was repeated 10 times.

In summary, there were $13\times 10=130$ runs on the MI dataset, each with $80\%\times(13\times160)=1,664$ clean training samples, $160$ poisoning samples, $20\%\times(13\times160)=416$ validation samples, and $160$ test samples. The mean ACCs and ASRs of these $13\times 10=130$ runs were computed and reported.

Among the eight subjects in the P300 dataset, one was randomly chosen to be the poisoning subject, and the remaining seven subjects to perform leave-one-subject-out cross-validation. We also performed under-sampling to the majority class to balance the training set. EEG trials, whose number equaled 10\% of the size of the training set, from the poisoning subject were randomly chosen to construct the poisoning set, all of which were labeled as `target'. The entire cross-validation process was repeated 10 times.

In summary, there were $7\times 10=70$ runs on the P300 dataset, each with $\sim$$5,800$ clean training samples, $\sim$$580$ poisoning samples, $\sim$$1,450$ validation samples, and $\sim$$3,300$ test samples. The mean ACCs and ASRs of these $70$ runs were computed and reported.

\subsection{Baseline performance.}

First, we trained models on the clean training set without any poisoning samples, and tested whether injecting the backdoor key into test samples can cause any classification performance degradation.

These baseline ACCs and ASRs of different classifiers on different datasets are shown in Table~\ref{tab:1}. The baseline ACCs were fairly high, considering the fact that they were evaluated on subjects different from those in the training set. The baseline ASRs were very small, indicating models that have not seen the poisoning data (the backdoor key) in training cannot be easily fooled by the backdoor key in test.

\begin{table*}[!h]  \center
\caption{\textbf{Baseline and NPP attack performance with different amplitude ratios.} NPP baseline: NPPs were used in test but not training; NPP attack: NPPs were used in both training and test. Low amp.: $10\%$/$50\%$/$0.1\%$ of the mean channel-wise standard deviation of the EEG amplitude for ERN/MI/P300; Middle amp.: $20\%$/$100\%$/$0.5\%$ for ERN/MI/P300; High amp.: $30\%$/$150\%$/$1\%$ for ERN/MI/P300.} \setlength{\tabcolsep}{0.7mm} \label{tab:1}
\begin{tabular}{c|c|cc|cc|cc|cc|cc|cc} \toprule
\multirow{3}{*}{Dataset} &\multirow{3}{*}{Model} &\multicolumn{6}{c}{NPP Baseline}  &\multicolumn{6}{|c}{NPP Attack}\\ \cline{3-14}
 & &\multicolumn{2}{|c}{Low amp.} &\multicolumn{2}{|c}{Middle amp.} &\multicolumn{2}{|c}{High amp.} &\multicolumn{2}{|c}{Low amp.} &\multicolumn{2}{|c}{Middle amp.} &\multicolumn{2}{|c}{High amp.} \\
	&     &ACC  &ASR  &ACC &ASR &ACC &ASR &ACC &ASR &ACC &ASR   &ACC  &ASR \\
\midrule
\multirow{3}{*}{ERN} &EEGNet &$0.661$ &$1.79\%$ &$0.655$ &$5.20\%$ &$0.648$ &$8.87\%$ &$0.673$ &$16.61\%$ &$0.655$ &$76.80\%$ &$0.653$ &$90.83\%$ \\
                    &DeepCNN &$0.638$ &$6.87\%$ &$0.643$ &$17.86\%$ &$0.640$ &$28.26\%$ &$0.635$ &$65.23\%$ &$0.638$ &$95.08\%$ &$0.635$ &$98.70\%$ \\
                    &xDAWN+LR &$0.651$ &$3.67\%$ &$0.648$ &$6.03\%$ &$0.651$ &$7.53\%$ &$0.676$ &$49.50\%$ &$0.665$ &$87.05\%$ &$0.663$ &$96.24\%$ \\
                    \midrule
\multirow{3}{*}{MI}  &EEGNet &$0.632$ &$2.23\%$ &$0.630$ &$2.26\%$ &$0.629$ &$4.98\%$ &$0.613$ &$14.42\%$ &$0.617$ &$56.29\%$ &$0.621$ &$86.56\%$\\
                     &DeepCNN &$0.614$ &$3.95\%$ &$0.608$ &$5.96\%$ &$0.612$ &$4.24\%$ &$0.598$ &$35.27\%$ &$0.593$ &$88.04\%$ &$0.589$ &$81.42\%$ \\
                     &CSP+LR &$0.544$ &$0.63\%$ &$0.542$ &$0.82\%$ &$0.544$ &$0.71\%$ &$0.512$ &$4.35\%$ &$0.504$ &$20.26\%$ &$0.501$ &$37.51\%$ \\
                    \midrule
\multirow{3}{*}{P300}&EEGNet &$0.667$ &$0.19\%$ &$0.667$ &$0.65\%$ &$0.668$ &$0.71\%$ &$0.563$ &$5.76\%$ &$0.664$ &$97.57\%$ &$0.665$ &$98.42\%$\\
                     &DeepCNN &$0.718$ &$0.27\%$ &$0.701$ &$0.77\%$ &$0.719$ &$0.62\%$ &$0.684$ &$0.93\%$ &$0.734$ &$97.44\%$ &$0.730$ &$98.38\%$ \\
                     &xDAWN+LR &$0.638$ &$1.33\%$ &$0.638$ &$6.59\%$ &$0.637$ &$8.11\%$ &$0.480$ &$96.40\%$ &$0.602$ &$98.96\%$ &$0.613$ &$99.45\%$ \\
\bottomrule
\end{tabular}
\end{table*}

\subsection{Attack performance.}

NPP backdoor keys with period $T=0.2$s, duty cycle $d=10\%$ and three different amplitudes were used for each dataset: $10\%/20\%/30\%$ of the mean channel-wise standard deviation of the EEG amplitude for the ERN dataset, $50\%/100\%/150\%$ for the MI dataset, and $0.1\%/0.5\%/1\%$ for the P300 dataset. These values were significantly different for different datasets, because the magnitudes of the raw EEG signals in different datasets varied a lot, possibly due to different hardware used and different experimental paradigms.

When the same NPP backdoor key was added to the poisoning samples and/or test samples, the attack performances are shown in the `NPP Attack' panel of Table~\ref{tab:1}. The ACCs were very close to those in the `NPP Baseline' panel, indicating that adding poisoning samples did not significantly change the classification accuracy, when the test samples did not contain the backdoor key. However, the ASRs in the `NPP Attack' panel were much higher than the corresponding baseline ASRs, indicating that these NPP backdoor attacks were very successful. Intuitively, the last two column of ASRs in the `NPP Attack' panel were higher than those in the first column, i.e., a larger NPP amplitude would lead to a higher ASR. Among different models, the traditional CSP+LR model seemed more resilient to the attacks.

\begin{figure*}[!ht]\centering
\includegraphics[width=.8\linewidth,clip]{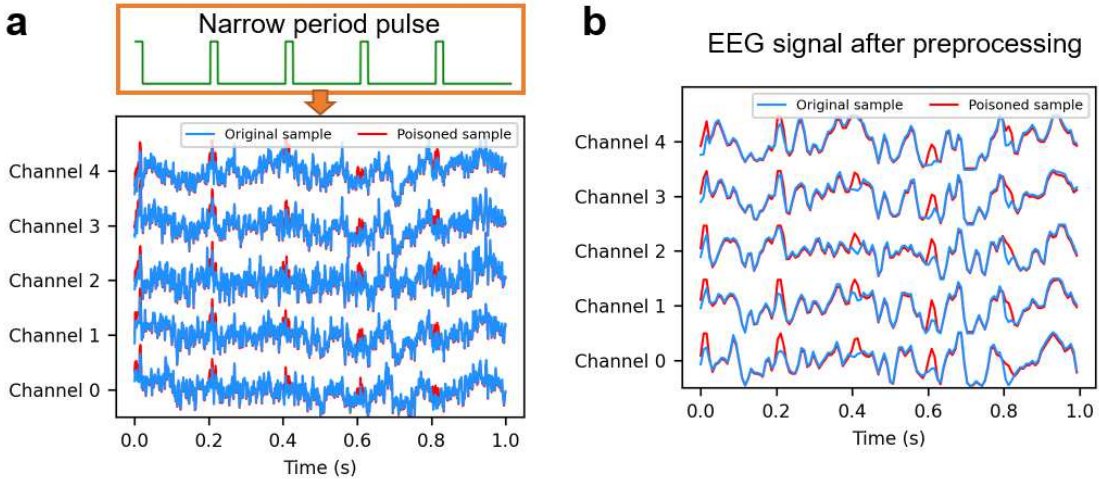}
\caption{\fontfamily{cmss}\selectfont  \textbf{EEG trial from the P300 dataset before (blue) and after (red) poisoning.} \textbf{a}. raw EEG trial without preprocessing. \textbf{b}. EEG trial after preprocessing.} \label{fig:2}
\end{figure*}

Figure~\ref{fig:2} shows an example of the same EEG trail from the P300 dataset before and after poisoning, with and without preprocessing (down-sampling and bandpass filtering), respectively. Examples for the ERN and MI datasets are shown in Supplementary Figures~1 and 2. The poisoned EEG looked like normal EEG, so the backdoor may not be easily detected. Additionally, preprocessing cannot remove the backdoor.

\subsection{Practical considerations.}

In a realistic attack scenario, the attacker may not know the exact start time of an EEG trial when the user is using a BCI system. As a result, the attacker cannot inject the backdoor key to a test sample exactly as he/she does in generating poisoning samples in training. So, a successful attack approach should not be sensitive to the start time of EEG trials.

To make the backdoor attacks more flexible and realistic, we used a random phase of NPP in $[0, 0.8]T$ ($T$ is the period of the NPP; see `Methods') for every poisoning sample. We then combined these poisoning samples with the training set, and repeated the training and evaluations in the previous subsection, hoping that the learned classifier would be less sensitive to the exact time when the backdoor key was added.

\begin{figure*}[!ht]\centering
\includegraphics[width=\linewidth,clip]{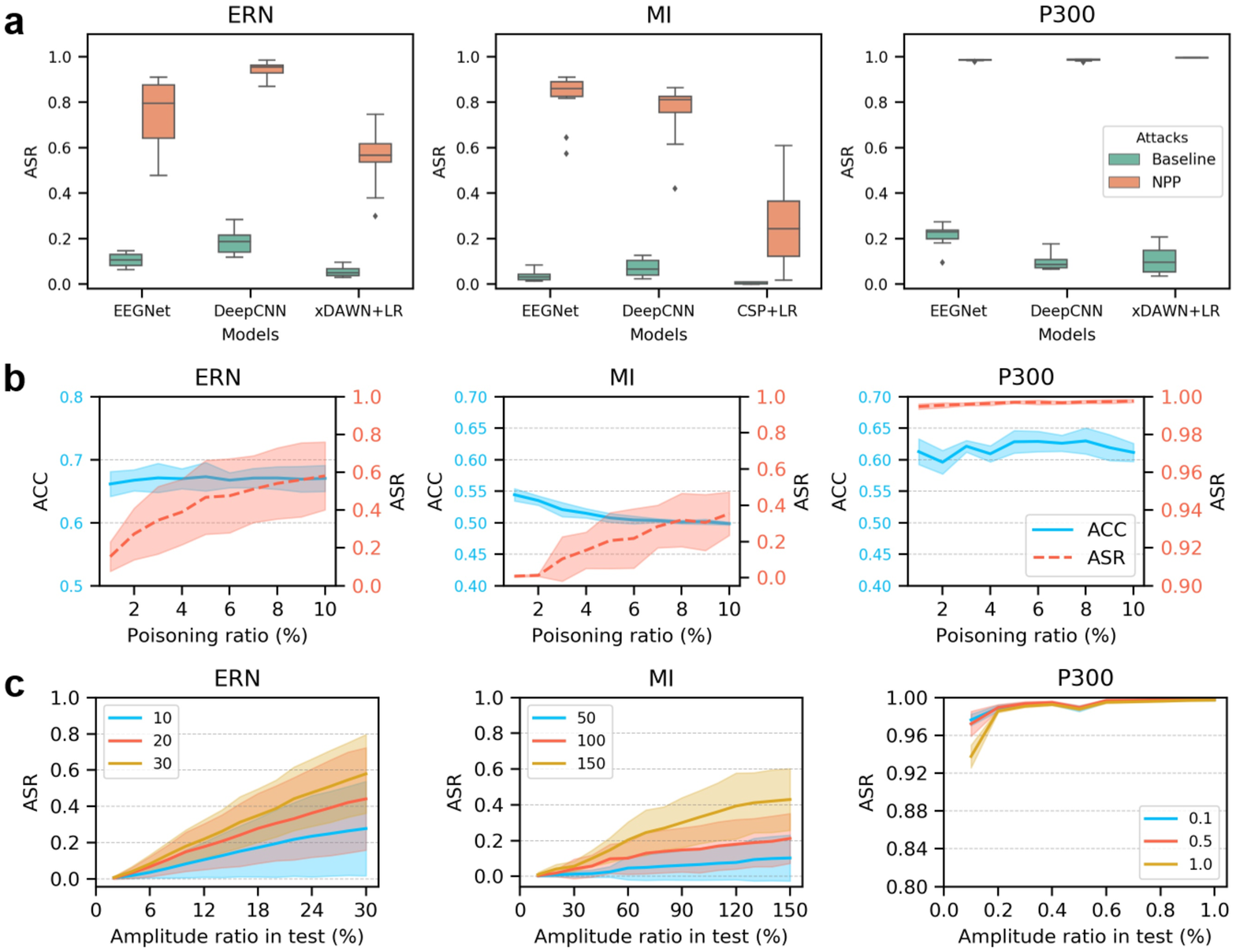}
\caption{\fontfamily{cmss}\selectfont \textbf{Poisoning attack results.} \textbf{a}. ASRs on the three datasets. \textbf{b}. Influence of the poisoning ratio to ACC and ASR for EEGNet. \textbf{c}. Influence of the amplitude ratio to ASR for EEGNet. The mean and standard deviations in \textbf{b} and \textbf{c} are computed from 10 repeats.} \label{fig:3}
\end{figure*}

The attack results of the approach on the three models are shown Figure~\ref{fig:3}a. NPP obtained much higher ASRs on different models than the baselines, indicating that the proposed NPP attack approach is insensitive to the start of EEG trials.

\subsection{Influence of the number of poisoning samples.}

Figure~\ref{fig:3}b shows the ACCs and ASRs of NPP attack to EEGNet when the poisoning ratio (the number of poisoning samples divided by the number of training samples) increased from $1\%$ to $10\%$. Results for DeepCNN and traditional models are shown in Supplementary Figures~3 and 4.

As the poisoning ratio increased, ACCs did not change much, whereas ASRs improved significantly. Only $6\%$ poisoning ratio on ERN and MI was enough to achieve an average ASR of $60\%$, and $1\%$ poisoning ratio on P300 achieved an average ASR of $80\%$. Compared with the large number of samples in the training set, the number of poisoning samples was very small, making the attacks very difficult to detect.

\subsection{Influence of the NPP amplitude.}

The NPP amplitude also affects ASRs. Figure~\ref{fig:3}c shows the ASRs of using NPPs with different amplitude ratios (the NPP amplitude divided by the mean channel-wise standard deviation of the EEG amplitude) in test for EEGNet. Results for other models are shown in Supplementary Figures~5 and 6. As the NPP amplitude ratio in test increased, the ASR also increased. The ASR also increased when larger NPPs were used in the poisoning set.

Interestingly, the NPP amplitude ratios may not need to match the amplitude ratios in training. For example, NPPs with amplitude ratio between $0.4\%$ and $1.0\%$ in test obtained similar ASRs on P300. In other words, the attacker does not need to know the exact NPP amplitude in poisoning, making the attack more practical.

\subsection{Influence of the NPP period and duty cycle.}

\begin{figure*}[h]\centering
\includegraphics[width=\textwidth,clip]{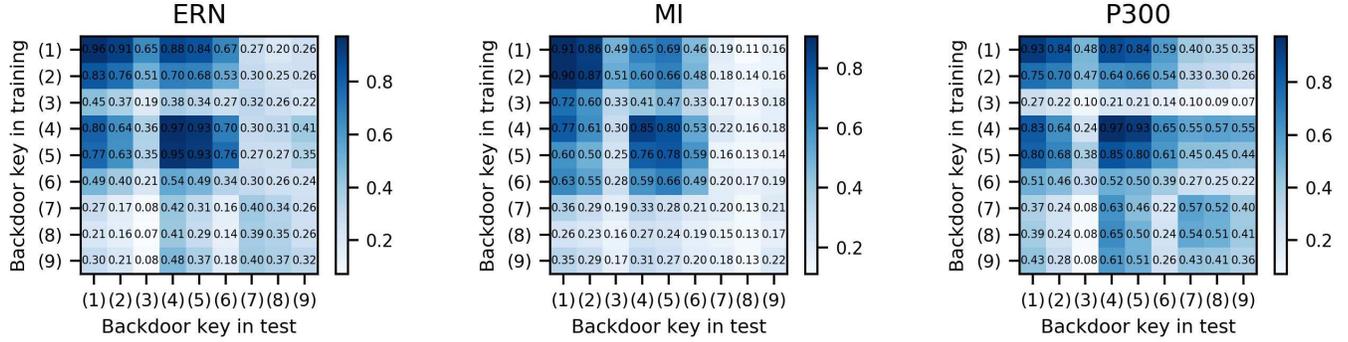}
\caption{\fontfamily{cmss}\selectfont \textbf{ASRs when NPPs with different periods and duty cycles are used in training and test of EEGNet.} The NPPs are: (1) $T=0.1$s and $d=15\%$; (2) $T=0.1$s and $d=10\%$; (3) $T=0.1$s and $d=5\%$; (4) $T=0.2$ s and $d=15\%$; (5) $T=0.2$s and $d=10\%$; (6) $T=0.2$ s and $d=5\%$; (7) $T=1$s and $d=15\%$; (8) $T=1$s and $d=10\%$; (9) $T=1$s and $d=5\%$. } \label{fig:4}
\end{figure*}

Figure~\ref{fig:4} shows the ASRs of using nine NPPs with different periods and duty cycles in training and test to attack EEGNet. Results on other models are shown in Supplementary Figures~7 and 8. Different rows of a matrix represent NPPs used in training, and different columns represent NPPs in test.

When NPPs were used in both training and test (the first six rows and six columns in Figure~\ref{fig:4}), high ASRs can be achieved, no matter whether the NPPs in training and test matched exactly or not, indicating that NPP attacks are also resilient to the NPP period and duty cycle. However, ASRs in the last three rows and three columns on ERN and MI dataset (Supplementary Figures~7 and 8) were relatively low, suggesting that the NPP parameters may impact ASRs in different BCI paradigms.

\subsection{Accommodate re-referencing.}

We have demonstrated the effectiveness and robustness of NPP attacks, without considering channel re-referencing\cite{Qin2010}, which may have some impact on the attack performance. For example, if we add identical NPPs to all EEG channels, then an average re-referencing\cite{Qin2010} would remove them completely, and hence the attack cannot be performed.

There are different solutions to this problem. If the attacker knows exactly the reference channel, e.g., Cz or mastoid, then NPPs can be added only to that channel. After referencing, NPP negations will be introduced to all other channels.

In practice, the attacker may not know what referencing approach and channels are used by the BCI system, so a more flexible solution is to add NPPs to a subset of channels. If average re-referencing is not performed, then NPPs in these channels are kept; otherwise, the NPP magnitudes in these channels are reduced but not completely removed.

Table~\ref{tab:2} shows the attack performance when NPPs were added to $10\%/20\%/30\%$ randomly selected EEG channels. The ASRs were comparable with or even higher than those of adding NPPs to all channels, suggesting that the attacker can add NPPs to a subset of channels to accommodate referencing.

\begin{table*}[!h]  \center
\caption{\textbf{Attack performance using different number of EEG channels}} \setlength{\tabcolsep}{2.5mm} \label{tab:2}
\begin{tabular}{c|c|cc|cc|cc|cc} \toprule
\multirow{3}{*}{Dataset} &\multirow{3}{*}{Model} &\multicolumn{8}{|c}{Number of EEG channels} \\ \cline{3-10}
 & &\multicolumn{2}{|c}{All channels} &\multicolumn{2}{|c}{$30\%$ channels} &\multicolumn{2}{|c}{$20\%$ channels} &\multicolumn{2}{|c}{$10\%$ channels} \\
	&  &ACC &ASR &ACC &ASR &ACC &ASR &ACC &ASR  \\ \midrule
\multirow{3}{*}{ERN} &EEGNet &$0.658$ &$75.45\%$ &$0.637$ &$95.63\%$ &$0.649$ &$94.78\%$ &$0.649$ &$83.21\%$ \\
                    &DeepCNN &$0.609$ &$94.34\%$ &$0.604$ &$97.24\%$ &$0.619$ &$95.26\%$ &$0.620$ &$85.54\%$ \\
                    &xDAWN+LR &$0.680$ &$56.06\%$ &$0.627$ &$100\%$ &$0.634$ &$100\%$ &$0.646$ &$100\%$ \\
                    \midrule
\multirow{3}{*}{MI}  &EEGNet &$0.621$ &$81.95\%$ &$0.627$ &$92.25\%$ &$0.632$ &$93.68\%$ &$0.629$ &$92.59\%$ \\
                     &DeepCNN &$0.603$ &$75.36\%$ &$0.601$ &$92.71\%$ &$0.605$ &$90.89\%$ &$0.614$ &$91.92\%$\\
                     &CSP+LR &$0.504$ &$25.75\%$ &$0.512$ &$55.06\%$ &$0.502$ &$63.40\%$ &$0.508$ &$27.46\%$ \\
                    \midrule
\multirow{3}{*}{P300}&EEGNet &$0.668$ &$98.57\%$ &$0.666$ &$97.42\%$ &$0.621$ &$95.31\%$ &$0.628$ &$86.91\%$\\
                     &DeepCNN &$0.746$ &$98.65\%$ &$0.743$ &$90.83\%$ &$0.714$ &$72.57\%$ &$0.704$ &$57.31\%$\\
                     &xDAWN+LR &$0.610$ &$99.70\%$ &$0.610$ &$99.98\%$ &$0.596$ &$99.70\%$ &$0.601$ &$99.96\%$ \\
\bottomrule
\end{tabular}
\end{table*}

\section*{Discussion}

Adversarial attacks to EEG-based BCIs have been explored in our previous studies\cite{Zhang2019,Meng2019,drwuALBCI2019,drwuNSR2020}. All of them were evasion attacks. These approaches are theoretically important, but very difficult to implement in practice. They all need to inject a jamming module between EEG preprocessing and machine learning, to add the adversarial perturbation to a normal EEG trial. It's difficult to implement in a real-world BCI system, in which EEG preprocessing and machine learning may be integrated. To generate or add the adversarial perturbation, the attacker also needs to know a lot of information about the target EEG trial, e.g., the start time is needed to align it with the adversarial perturbation, but it is very difficult to know this. Furthermore, the adversarial perturbations generated by these attack approaches are very complex for a real-world BCI system to realize, e.g., different channels need to have different adversarial perturbations, which are very challenging to add.

Compared with previous approaches, the NPP backdoor attack approach proposed in this article is much easier to implement, and hence represents a more significant security concern to EEG-based BCI systems.

Our future research will consider hardware implementation of the attack approach in a real-world BCI system, and more importantly, strategies to defend against such attacks, as the ultimate goal of our research is to increase the security of BCI systems, instead of damaging them.

\small
\section*{Methods}

Assume the model designer has a labeled training set $\mathcal{D} = \{(\bm{x}_i,y_i)\}_{i=1}^N$ with $N$ samples, which are not known to the attacker. The attacker also has some normal samples $\{\bm{a}_j\}_{j=1}^M$, where $\bm{a}$ and $\bm{x}$ have the same format, and usually $M\ll N$.

The attacker aims to design a backdoor key $\bm{k}$, and a function $g(\bm{a}_{j},\bm{k})$ which adds $\bm{k}$ to $\bm{a}_{j}$ to form a poisoning sample. The attacker then adds $\{(g(\bm{a}_j,\bm{k}),y)\}_{j=1}^M$ to $\mathcal{D}$, where $y$ is the target class specified by the attacker, i.e., he/she wants any test sample with the backdoor $\bm{k}$ to be classified into class $y$.

The model designer trains a classifier on the poisoned training set $\mathcal{D}'=\mathcal{D}\cup \{(g(\bm{a}_j,\bm{k}),y)\}_{j=1}^M$, using whatever classifier he/she wants, e.g., traditional machine learning or deep learning. The backdoor $\bm{k}$ is automatically embedded into the model.

During the attack, the attacker can add $\bm{k}$ to any benign test sample $\bm{x}$ to open the backdoor and force the classifier to classify $\bm{x}$ to the target class $y$. When $\bm{k}$ is not added, the BCI system just operates normally.

\subsection{Narrow period pulse (NPP).}

NPP is a type of signal that can be easily generated. A continuous NPP is determined by a period $T$, a duty cycle $d$ (the ratio between the pulse duration and the period), and an amplitude $a$,
\begin{align*}
	\mathcal{N}_c(t) = \left\{
             \begin{array}{ll}
             a, &  nT\leq t < nT+dT \\
             0, &  nT+dT\leq t < (n+1)T
             \end{array}
             \right.
\end{align*}
An example of continuous NPP is shown at the top of Figure~\ref{fig:3}.

A discrete NPP with sampling rate $f_s$ can be expressed as
\begin{align*}
	\mathcal{N}_d(i) = \left\{
             \begin{array}{ll}
             a, &  nTf_s\leq i < (n+d)Tf_s \\
             0, &  (n+d)Tf_s\leq i < (n+1)Tf_s
             \end{array}
             \right.
\end{align*}
This NPP was used as $\bm{k}$, and $g(\bm{a}_j,\bm{k})=\bm{a}_j+\bm{k}$ in obtaining the results in Table~\ref{tab:1}.

A discrete NPP with a random phase $\phi\in[0,T]$ is expressed as
\begin{align*}
	\mathcal{N}_d(i) = \left\{
             \begin{array}{ll}
             0, &  nTf_s\leq i < (nT+\phi)f_s \\
             a, &  (nT+\phi)f_s\leq i < (nT+dT+\phi)f_s \\
             0, &  (nT+dT+\phi)f_s\leq i < (n+1)Tf_s
             \end{array}
             \right.
\end{align*}
This NPP was used as $\bm{k}$, and $g(\bm{a}_j,\bm{k})=\bm{a}_j+\bm{k}$ in obtaining the results in Figure~\ref{fig:3}.

\newpage
\section*{References}\ \vspace*{10pt}

\end{document}